\documentclass{emulateapj}
\usepackage{apjfonts}
\usepackage{graphicx}
\pdfoutput=1

\newcommand\lsim{\mathrel{\rlap{\lower4pt\hbox{\hskip1pt$\sim$}}
        \raise1pt\hbox{$<$}}}
\newcommand\gsim{\mathrel{\rlap{\lower4pt\hbox{\hskip1pt$\sim$}}
        \raise1pt\hbox{$>$}}}
\newcommand\propsim{\mathrel{\rlap{\lower4pt\hbox{\hskip1pt$\sim$}}
        \raise1pt\hbox{$\propto$}}}

\newcommand{\msun}{{\rm M_{\odot}}}

\begin{document}

\title{Supermassive Black Hole Formation at High Redshifts Through \\a Primordial Magnetic Field}

\author{Shiv Sethi$^{1,3}$, Zolt\'{a}n Haiman$^{2}$ \& Kanhaiya  Pandey$^1$}
\affiliation{
$^1$Raman Research Institute, C. V. Raman Avenue, Sadashivanagar, Bengalooru, 560 080, India; (sethi,kanhaiya)@rri.res.in\\
$^2$Department of Astronomy, Columbia University, 550 West 120th Street, New York, NY 10027, USA; zoltan@astro.columbia.edu \\
$^3$ McWilliams Center for Cosmology, Department of Physics, Carnegie Mellon University, 5000 Forbes Ave., Pittsburgh, PA 15213, USA \\
}
\vspace{+0.4cm}

\begin{abstract}
  It has been proposed that primordial gas in early dark matter halos,
  with virial temperatures $T_{\rm vir}\gsim 10^4$K, can avoid
  fragmentation and undergo rapid collapse, possibly resulting in a
  supermassive black hole (SMBH).  This requires the gas to avoid
  cooling and to remain at temperatures near $T\sim 10^4$ K.  We show
  that this condition can be satisfied in the presence of a
  sufficiently strong primordial magnetic field, which heats the
  collapsing gas via ambipolar diffusion.  If the field has a strength
  above $\mid B\mid\gsim 3.6$ (comoving) nG, the collapsing gas is
  kept warm ($T\sim 10^4$ K) until it reaches the critical density
  $n_{\rm crit}\approx10^3 {\rm cm^{-3}}$ at which the
  roto--vibrational states of ${\rm H_2}$ approach local thermodynamic
  equilibrium.  ${\rm H_2}$--cooling then remains inefficient, and the
  gas temperature stays near $\sim 10^4$K, even as it continues to
  collapse to higher densities. The critical magnetic field strength
  required to permanently suppress ${\rm H_2}$--cooling is somewhat
  higher than upper limit of $\sim 2$nG from the cosmic microwave
  background (CMB). However, it can be realized in the rare
  $\gsim(2-3)\sigma$ regions of the spatially fluctuating $B$--field;
  these regions contain a sufficient number of halos to account for
  the $z\approx6$ quasar BHs.
\end{abstract}

\vspace{\baselineskip}


\vspace{\baselineskip}

\section{Introduction}

The discovery of very bright quasars, with luminosities $\ge
10^{47}~{\rm erg~s^{-1}}$, at redshift $z\simeq 6$ in the Sloan
Digital Sky Survey (SDSS) suggests that some SMBHs as massive as a
few$\times10^9~\msun$ already existed when the universe was less than
1 Gyr old (see, e.g., Fan 2006 for a review).  The presence of these
SMBHs presents a puzzle.  Metal--free stars, with masses $\sim
100~\msun$, are expected to form at redshifts as high as $z\gsim 25$
(Abel et al. 2002; Bromm et al. 2002; Yoshida et al. 2008), and leave
behind remnant BHs with similar masses (Heger et al. 2003).  However,
the natural time-scale, i.e. the Eddington time, for growing these
seed BHs by $\gsim 7$ orders of magnitude in mass is comparable to the
age of the universe (e.g. Haiman \& Loeb 2001). This makes it
difficult to reach $10^9~\msun$ without a phase of rapid
(super--Eddington) accretion, unless a list of optimistic assumptions
are made in hierarchical merger models, in which multiple seed BHs are
allowed to grow without interruption, and to combine into a single
SMBH (Haiman 2004; Yoo \& Miralda-Escud\'{e} 2004; Bromley et
al. 2004; Shapiro 2005, Volonteri \& Rees 2006; Li et al. 2007; Tanaka
\& Haiman 2009).

An alternative class of explanations involves rapid gas accretion or
collapse. In this family of models, primordial gas collapses rapidly
into a SMBH as massive as $10^4-10^6~\msun$ (Oh \& Haiman 2002; Bromm
\& Loeb 2003; Koushiappas et al. 2004; Lodato \& Natarajan 2006;
Spaans \& Silk 2006; Begelman et al. 2006; Volonteri et al. 2008; Wise
\& Abel 2008; Regan \& Haehnelt 2009; Shang et al. 2010), possibly by
accreting onto a pre--existing smaller seed BH (Volonteri \& Rees
2005), or going through the intermediate state of a very massive star
(Bromm \& Loeb 2003), a dense stellar cluster (Omukai et al. 2008), or
a ``quasistar'' (Begelman et al. 2008).  These so--called ``direct
collapse'' models involve metal--free gas in relatively massive
($\gsim 10^8~\msun$) dark matter halos at redshift $z\gsim 10$, with
virial temperatures $T_{\rm vir}\gsim 10^4$K.  The gas that cools and
collapses in these halos must avoid fragmentation, shed angular
momentum efficiently, and collapse rapidly.  These conditions are
unlikely to be met, unless the gas remains ``warm'', i.e. at
temperatures $T_{\rm vir}\sim 10^4$K.  In particular, in recent
numerical simulations, Shang et al. (2010) found that the gas in such
halos, when collapsing in isolation, forms ${\rm H_2}$ efficiently,
and cools to temperatures of $T \sim 300$ K.  Although no
fragmentation was seen, the gas could ultimately fragment on smaller
scales that have not yet been resolved (e.g. Turk et al. 2009).  More
importantly, even if fragmentation was avoided, the cold gas flows
inward at low velocities, near the sound speed of $\sim 2-3~{\rm
km~s^{-1}}$, with a correspondingly low accretion rate of $\sim
0.01~{\rm M_\odot~yr^{-1}}$. This results in conditions nearly
identical to those in the cores of lower-mass minihalos; extensive
ultra--high resolution simulations have concluded that the gas then
forms a single $\sim 100~{\rm M_\odot}$ star (Abel et al. 2002; Bromm
et al. 2002; Yoshida et al. 2008) or perhaps a massive binary (Turk et
al. 2009), rather than a supermassive star or BH.

${\rm H_2}$--cooling in early galaxies may be avoided when the gas is
exposed to an intense UV flux $J$, either directly photo--dissociating
${\rm H_2}$ (in the Lyman--Werner bands near a photon energy of $\sim
12$eV) or photo--dissociating the intermediary ${\rm H^-}$ (at photon
energies $\gsim 0.76$eV).  Requiring the photo-dissociation timescale,
$t_{\rm diss}\propto J^{-1}$, to be shorter than the ${\rm
H_2}$--formation timescale, $t_{\rm form}\propto \rho^{-1}$,
generically yields a critical flux that increases linearly with
density, $J^{\rm crit} \propto \rho$.  Since the gas in halos with
$T_{\rm vir}\gsim 10^4$K can cool via atomic Lyman $\alpha$ radiation
and loose pressure support, it inevitably collapses further. As a
result, in these halos, the critical flux is high, $J^{\rm
crit}\approx10^{2}-10^{5}$ (depending on the assumed spectral shape;
Shang et al. 2010; see also Omukai 2001 and Bromm \& Loeb 2003 who
found similar, but somewhat higher values), which exceeds the expected
level of the cosmic UV background at high redshifts.  Only a small
subset of all $T_{\rm vir}\gsim 10^4$K halos, which have unusually
close and bright neighbors, may see a sufficiently high flux (Dijkstra
et al. 2008). In order to avoid fragmentation, the gas in these halos
must also remain essentially free of any metals and dust (Omukai et
al. 2008), which may be incompatible with the presence of such nearby
luminous galaxies.

In this paper, we consider a different possibility to keep the gas
warm, relying on heating by a primordial magnetic field.  Several
mechanisms have been proposed to produce a global primordial magnetic
field (PMF), with a field strength of order 1 (comoving) nG, during
inflation and/or during various phase transitions in the early
universe (see, e.g. Widrow 2002 and references therein).  If present,
the PMF can be strongly amplified by flux--freezing inside a
collapsing primordial gas, affecting ${\rm H_2}$--formation and
cooling. Sethi et al. (2008; hereafter S08) have shown that 0.2-2 nG
fields can significantly enhance the ${\rm H_2}$ fraction during the
early stages of collapse. Schleicher et al. (2009) found similar
results, and emphasized that at the high densities ($n\gsim 10^8~{\rm
cm^{-3}}$) corresponding to later stages of the collapse, magnetic
heating from 0.1-1nG fields results in significantly elevated
temperatures.

In this paper, we consider field strengths higher than those in the
two previous studies by S08 and Schleicher et al. (2009).  Our main
result is that, in analogy with the UV--irradiation, there exists a
critical magnetic field, which leads to a bifurcation of behaviors.
If the PMF has a strength above $B>B_{\rm crit}\approx3.5$ nG, then
the collapsing gas is kept warm ($T\sim 10^4$ K) until it reaches the
critical density $n_{\rm crit}\approx10^3 {\rm cm^{-3}}$ at which the
roto--vibrational states of ${\rm H_2}$ approach local thermodynamic
equilibrium.  ${\rm H_2}$--cooling then remains inefficient, and the
temperature stays near $\sim 10^4$K, even as the gas collapses
further.  On the other hand if $B<B_{\rm crit}$, ${\rm H_2}$--cooling
is delayed, but the gas eventually cools down below $\sim 1000$K.  The
critical magnetic field strength we find is a factor of $\sim$two
higher than the existing upper limit from the CMB anisotropies
(Yamazaki 2010; see more discussion in \S~5 below).  However, it can
be realized in the rare $\gsim(2-3)\sigma$ regions of the spatially
fluctuating $B$--field. As we argue, the abundance of halos located in
these high--field regions is sufficient to explain the number of the
$z\approx6$ quasars observed in the SDSS.

\section{Chemistry and Thermal Evolution of Collapsing Primordial Gas}

The thermal and ionization history of gas collapsing into a
high--redshift halo, in the presence of a primordial magnetic field,
is described in detail in S08. The dissipation of the magnetic field
owing to ambipolar diffusion and to decaying turbulence in the post
recombination era can substantially alter the ionization fraction and
temperature of the gas, even beginning before the halo collapse (Sethi
\& Subramanian 2005 [hereafter SS05]; Yamazaki et~al. 2006; S08;
Schleicher et~al. 2009).

Here we extend the analysis of S08 up to a higher particle density in
the collapsing halo, and explore higher magnetic field strengths. In
addition, we made the following changes: (a) we track the density
evolution of gas in a collapsing halo with the model of Dekel and
Birnboim (2006; this affects the compressional heating rate), (b) we
updated the $\rm H_2$ chemistry network; specifically, we use the
recent compilation in Shang et~al. (2010; the most significant change
is an increase in the collisional ${\rm H_2}$--dissociation rate).

The evolution of the ionization fraction ($x_e$), magnetic field
energy density ($E_B$), temperature ($T$), and ${\rm H_2}$ molecule
fraction ($x_{\rm H_2}$) are described by the equations
\begin{eqnarray}
\dot x_e  =& &    \left [ \beta_e (1-x_e)\exp\left(-h\nu_\alpha/(k_BT_{\rm cbr})\right )-\alpha_e n_b x_e^2 \right ] C + 
 \nonumber \\
  & & +\, \gamma_e n_b (1-x_e)x_e \\
{dE_B \over dt}&  = & {4 \over 3} {\dot \rho \over \rho}  - \left({dE_B \over dt}\right)_{\rm turb}  - \left({dE_B \over dt}\right)_{\rm ambi} \\
{dT \over dt}&=&{2 \over 3} {\dot{n_b} \over n_b} T+k_{iC} x_e (T_{\rm cbr} -T)
+{2 \over 3 n_b k_B} (L_{\rm heat} -L_{\rm cool}) \,, \\ 
{d x_{\rm \scriptscriptstyle  H_2} \over dt}&=&k_m n_b x_e (1-x_e-2 x_{\rm \scriptscriptstyle  H_2}) - k_{\rm des} n_b  x_{\rm \scriptscriptstyle  H_2} \,.
\end{eqnarray}
The symbols here have their usual meaning; further details and
relevant processes are discussed in S08. We list here only the
processes that have been added or updated.  Most of these processes
relate to the formation and destruction of $\rm H_2$.  The net rate of
formation of $\rm H_2$ through the $\rm H^-$ channel is given by:
\begin{equation}
k_m= {k_9 k_{10}  x_{HI} n_b \over k_{10} x_{HI} n_b + k_\gamma +
(k_{13}+k_{21}) x_p n_b +k_{19}x_e n_b +k_{20} x_{HI} n_b} \,.
\end{equation}
The notation of the reaction rates follows the appendix of Shang
et~al. (2010), except that $k_\gamma$ is the rate of destruction of
$H^-$ by CMB photons, which, in our case, is important before the
collapse regime (eq.~8 in S08).  The net destruction rate of $\rm
H_2$, $k_{\rm des}$, is:
\begin{equation}
k_{\rm des} = k_{15}x_{HI}+k_{17} x_p+k_{18}x_e.
\end{equation}
The dominant reaction rates for the range of ionization fractions,
densities, and temperatures we obtain in the entire range of our
models are: $k_9$, $k_{10}$, $k_{20}$, $k_{18}$, and
$k_{15}$\footnote{We note here that the rate $k_{14}$, assumed to be
zero here, and $k_{18}$ could become more important if we used the
rates as given by Schleicher et~al. (2009) and Capitelli
et~al. (2007). If these rates are used, then the critical value of
$B_0$ needed to destroy $\rm H_2$ in the collapsing halos decreases to
$\simeq 3 \, \rm nG$ (cf. Figure~1 below).}

The cooling processes that dominate $L_{\rm cool}$ in primordial gas
in the density and temperature range we consider are: (a) Compton
cooling $k_{iC}$ (eq.~15 in S08), (b) atomic H cooling (eq.~16 in
S08), and (c) ${\rm H_2}$ molecular cooling (Galli \& Palla 1998).
The heating rate $L_{\rm heat}$ is given by the magnetic field decay
owing to decaying turbulence $(dE_{\rm B}/dt)_{\rm turb}$ and
ambipolar diffusion $(dE_{\rm B}/dt)_{\rm ambi}$.  In practice, we
find that ambipolar diffusion always dominates in our case; the
dissipation rate is given for this process by (Cowling 1956; Shu 1992;
SS05):
\begin{equation}
\left({dE_B \over dt}\right)_{\rm ambi} = {7\rho_n f(t) \over 48 \pi^2 \gamma \rho_b^2 \rho_i}   \int dk_1 
\int dk_2 M(k_1)  M(k_2) k_1^2 k_2^4.
\label{eq:amdif}
\end{equation}
All quantities in eq.~(\ref{eq:amdif}) are expressed at redshift
$z=0$. The time--dependence of the decay rate is given by $f(t) =
(1+z)^4$ during the pre-collapse stage, and $f(t) \propto \rho^{-4/3}$
during the collapse phase.  Here $M(k) = Ak^n$ is the power spectrum
of the tangled magnetic field, with a large-$k$ cut--off at $k =
k_{\rm max} \simeq 235 (1 \, {\rm nG}/B_0) \, \rm (comoving)
Mpc^{-1}$; $k_{\rm max}$ is determined by the effects of damping by
radiative viscosity during the pre-recombination era (Jedamzik,
Katalini{\' c}, \& Olinto 1998, Subramanian \& Barrow 1998a; see SS05
for further details). $B_0$, referred to as the magnetic field
strength, is defined as the r.m.s. value at $k = 1 \, \rm Mpc^{-1}$.
Throughout this paper, we use $n = -2.9$ (for justification and
further discussion see SS05, S08 and references therein).  Unless
specified otherwise, the time--evolution of the magnetic field is
assumed to be given by flux-freezing, which implies a power--law
dependence $B\propto\rho^{\alpha}$ on the gas density with
$\alpha=2/3$. In practice, this scaling may be less steep; below we
will explore how our results change for different values of $\alpha$.

Another important scale is the (comoving) magnetic Jeans scale, $k_J
\simeq 15 (1 \, {\rm nG}/B_0) \, \rm Mpc^{-1}$ (see e.g.  Eq.~6 in
SS05 and references therein). This scale, along with the thermal Jeans
scale determine the condition that allow a halo to be gravitationally
unstable.  In our case, the magnetic Jeans scale is larger, and is
therefore more restrictive, for $B_0 \ge 1 \, \rm nG$ (S08, Figure~4).
Primordial magnetic fields can also induce the formation of first
structures in the universe (Wasserman 1978, Kim et~~al. 1996, Gopal
and Sethi 2003, Subramanian and Barrow 1998a, SS05). The total (dark
matter + gas) mass of the first structures, determined by the magnetic
Jeans length, is $M \simeq 5 \times 10^{8} (B_0/1 \, {\rm nG})^3 \,
\rm M_\odot$ (e.g. Figure~7 in SS05).

\section{Results and Discussion}

We show in Figure~1, 2, and 3 the evolution of the temperature, the
$\rm H_2$ fraction (defined as the ratio of the ${\rm H_2}$ number
density to the total hydrogen number density, $n_{\rm
\scriptscriptstyle H2}/n_{ \rm \scriptscriptstyle H}$), and the
ionized fraction for a single halo from $z \simeq 800$ (corresponding
to the initial number density $n \simeq 100 \, \rm cm^{-3}$ on the
left of Figures~1, 2, and 3), down to a maximum density of $n \simeq
10^6 \, \rm cm^{-3}$ in the collapsed halo. Note that the evolution on
these figures is monotonically to the right: the x--axis shows the
density decreasing to the right (until the turnaround redshift), and
then increasing again as the halo collapses.

The figures show the interplay between several physical effects.
First, the magnetic field decay directly increases the
temperature. This increases the collisional destruction rate of ${\rm
H2}$, but it also increases the electron fraction (owing to more rapid
collisional ionization).  The larger electron fraction then tends to
increase the molecular hydrogen fraction, competing with the effect of
the increased collisional ${\rm H_2}$--dissociation.  The molecular
hydrogen cooling rate depends on the temperature directly, and also on
the molecular hydrogen fraction.  As the temperature reaches $\gsim
8,000$K, atomic cooling dominates, which, again, is governed by the
ionized fraction.

Figures~1, 2, and~3 should be viewed together to appreciate the net
outcome of these effects on the thermal evolution, for different
specific values of the magnetic field.  A higher magnetic field
strength generally results in more rapid heating, and therefore a
higher temperature, at least until halo collapse begins. This, in turn
results in higher ionized fractions.  However, the thermal evolution
is more complicated in the collapsing stage, once molecular hydrogen
becomes the dominant coolant.

Prior to turnaround, the gas temperature increases roughly
monotonically with the strength of the magnetic field, up to $10^4$K.
However, since a higher temperature also means a higher ionized
fraction (Fig.~2), molecular hydrogen forms at a faster rate.  Once
the molecular hydrogen becomes the dominant coolant, at $n \simeq 1 \,
\rm cm^{-3}$, the halo with a higher magnetic field can cool below the
less magnetized halo.

The most interesting case corresponds to $B_0 = 4 \, \rm nG$ (shown by
the thick solid black curves). In this case, the magnetic field
dissipation rate is high enough to prevent the formation of molecular
hydrogen in the collapsing halo. As seen in Figure~2, the molecular
hydrogen starts getting destroyed in the collapsing halo in this
case. The direct impact of this predicament on the thermal state of
the gas is that the gas fails to cool via ${\rm H_2}$ -- the cooling
remains dominated by the line excitation of the atomic hydrogen.

Our results are in reasonable agreement with the results of Schleicher
et~al. (2010) for $B_0 \le 1 \, \rm nG$, the range of magnetic field
strengths they considered. We note that the molecular hydrogen
fractions, past the turn--around epoch, are lower by more than an
order of magnitude compared to the results of S08.  This is in line
with the observation of Schleicher et~al. (2009) that S08 had
underestimated the destruction rate of $\rm H_2$.  The thermal
evolution of the gas in the zero magnetic field case ($B_0 \le 10^{-3}
\, \rm nG$ in the figures) also agrees with the results of the
one-zone models studied by Omukai (2001).

The most important new result of our analysis is the thermal state of
the gas owing to the destruction of the $\rm H_2$ in the collapsing
halo for $B \ge 4 \, \rm nG$. Omukai (2001) analyzed the destruction
of $\rm H_2$ in the collapsing halo in the presence of background UV
flux. That analysis suggested that the $\rm H_2$ formation can be
prevented if the halo could be kept at a temperature $\simeq 10^4$ up
to a critical density $n \simeq 2000$. At higher densities, collisions
with hydrogen atoms is more effective in destroying $\rm H_2$ from
higher vibration levels, because the relative occupation probability
of vibration states of the $\rm H_2$ molecular approach thermal
equilibrium at these densities (the rate $k_{15}$ in the discussion in
section~2, Martin et~al. 1996). In the analysis of Omukai (2001), the
reaction rates are also affected by the presence of a background UV
field. Our analysis shows that the same result (i.e. lack of any ${\rm
H_2}$ cooling) can also be obtained by the dissipation of tangled
magnetic fields.

In Figures~4 and 5, we show the heating and cooling rates during the
collapse regime, for two values of the magnetic field, 1 and 4~nG,
respectively. Apart from a brief initial period, when atomic cooling
roughly balances adiabatic heating, the dominant heating and cooling
processes in the $B_0 = 1 \, \rm nG$ case, in the collapse stage, are
magnetic heating and $\rm H_2$ cooling, respectively.  $\rm H_2$
cooling quickly becomes more important than magnetic heating,
resulting in a rapid temperature drop at densities near $n\approx
0.1-0.2~{\rm cm^{-3}}$ (see Fig.~1).  As the halo collapses further,
the gas begins to recombine and the ionized fraction decreases (see
Fig.~3), the magnetic field dissipation rate due to ambipolar
diffusion increases as $\propto \rho_b^{4/3}/\rho_i$, while the $\rm
H_2$ cooling rate grows as $x_{\rm \scriptscriptstyle
H_2}\rho_b^2$. As a result, the magnetic heating catches up with $\rm
H_2$ cooling, resulting in a nearly constant temperature (see Fig.~1)
-- however, this occurs only after the collapse has proceeded beyond
the critical density $n\approx 10^{3}~{\rm cm^{-3}}$.  Atomic cooling
or adiabatic heating do not play an important role in the thermal
evolution for this strength of the magnetic field.

In contrast, for $B_0 \simeq 4 \, \rm nG$, as shown in Figure 5, the
magnetic heating roughly balances atomic HI cooling during the
collapse stage, resulting in a nearly constant temperature $T\approx
10^4$K throughout the entire evolution of the contracting gas.  The
magnetic heating gives rise to a higher ionization fraction (Fig.~3),
and therefore aids the formation of $\rm H_2$.  However, the high
temperature of the gas causes the $\rm H_2$ to get destroyed (Fig.~2;
this effect is also seen for $B_0 \simeq 1 \, \rm nG$ just before the
onset of the halo collapse).  The net result for this high value of
the magnetic field is that $\rm H_2$ cannot form fast enough to ever
become an important coolant.  The halo remains at a temperature
$\simeq 10^4 \, \rm K$ up to the critical density; as a result, the
$\rm H_2$ fraction is strongly diminished for the subsequent evolution
of the halo.  By experimenting with several intermediate values of the
magnetic field, we have found that this clear--cut bifurcation in the
thermal evolution occurs at a critical magnetic field strength of
$B_{\rm crit}=3.6 \, \rm nG$.

As noted in \S~2 above, we adopted the model of Dekel and Birnboim
(2006) to track the density evolution of the collapsing halo.
However, we have checked the robustness of our results against a wide
range of collapse histories. Specifically, after the turn--around
stage, we artificially multiplied the rate of increase in the density,
relative to the Dekel \& Birnboim model, by a factor of 0.1 or 1.  We
have found that the weaker/stronger adiabatic heating delayed/advanced
the onset of the catastrophic ${\rm H_2}$ cooling (e.g. as seen at
$n\approx 10^{-0.8}{\rm cm^{-3}}$ in Fig.1) to higher/lower densities,
but the other qualitative features of our results were essentially
unchanged. In particular, a bifurcation of behaviors was still found,
with a critical density of $B_{\rm crit}=3.6\, \rm nG$.

Another uncertainty concerns our assumption of
flux--freezing. Schleicher et~al. (2009) note the possibility of the
breakdown of this approximation in a collapsing halo. If the field is
not sufficiently tangled, collapse can occur with little dissipation
in the direction of the field lines; the magnetic field might grow
less rapidly than our adopted $\rho^{2/3}$. Our computations in the
range $\alpha \simeq 0.55\hbox{--}0.6$ show that the critical magnetic
field required to prevent the halo from cooling increases to $B_0
\simeq 5\hbox{--} 7 \, \rm nG$.  This also leads to the interesting
possibility that the average $B_0 \simeq \hbox{a few} \, \rm nG$,
required to form SMBHs, might be detectable by the future CMB
experiments (e.g. Yamazaki et~al. 2010) and the $4\hbox{--}5\sigma$
fluctuations of the field might leave their trace in the formation of
SMBH. We hope to explore this possibility in the future.

At present, the best upper limits on the PMF are in the range
$2\hbox{--}3 \, \rm nG$ from CMB temperature and polarization
anisotropies and from early structure formation (Subramanian \& Barrow
1998b, 2002, Durrer, Ferreira \& Kahniashvili 2000, Seshadri \&
Subramanian 2001, Mack, Kahniashvili \& Kosowsky 2002, Lewis 2004,
Gopal \& Sethi 2005, Kahniashvili \& Ratra 2005, Giovannini \& Kunze
2008, Yamazaki et~al. 2008, Finelli, Paci \& Paoletti 2008). Recently,
Yamazaki et~al. (2010) obtained bounds on the strength and spectral
index of the magnetic field power spectrum: $B_0 \le 2 \, \rm nG$ and
$n \le -1.4$.  Our results cannot be compared directly to their
analysis as we use only a single value of spectral index ($n = -2.9$)
here. The upper limits from CMB on $B_0$ for this value of $n$ are
considerably weaker (Lewis 2004). We note that the effect we discuss
here might be pronounced for a larger $n$ for a given $B_0$. More
detailed analysis would be required to directly compare our results
with Yamazaki et~al. (2010). Our analysis suggests that for $B_0 \ge 4
\, \rm nG$, the fragmentation of collapsing halos could be prevented,
and therefore this could be considered an independent upper bound on
the value of $B_0$.

\section{The Mass of the Central Object}

The mass of the central object that forms in a collapsing
proto--galaxy can be approximated as follows: there exists a radius at
which the mass accretion time--scale $t_{\rm acc}$ equals the
Kelvin-Helmholtz time scale $t_{\rm KH}$ for a proto--star, with the
proto--stellar mass equal to the gas mass enclosed within this radius.
For metal--free gas, $t_{\rm KH}$ is approximately $10^5$ years, with
only a mild dependence on the proto--stellar mass (Schaerer 2002). The
expected mass of the central object then scales approximately as $M
\propto t_{\rm acc}^{-1}\propto c_s^3\propto T^{3/2}$ (see, e.g.,
Shang et al. 2010 for the last scalings with the sound speed and gas
temperature).  This implies that a stellar mass of $\sim 200 {\rm
M_\odot}$, expected for $T=300$K, can increase to $\approx4\times
10^{4}~{\rm M_\odot}$ when ${\rm H_2}$--cooling is inefficient and
$T\approx10^4$K (in their three--dimensional simulations, Shang et
al. find a somewhat still steeper scaling).  Our proposal here is that
a small fraction of halos at $z=10-15$, which contain pristine,
metal--free gas when they collapse, and which reside in regions of an
unusually high initial seed magnetic field, may produce a SMBH with a
mass of up to $\sim 10^{4-5}~{\rm M_\odot}$.  The time available
between $z=6$ and $z=10-15$ is $\approx(4-6)\times 10^8$ yrs, allowing
for a further growth in mass by a factor of $\approx(2\times10^4) -
(3\times 10^6)$ at the e--folding time of $4\times 10^7~{\rm yr}$,
(corresponding to Eddington--limited growth at the radiative
efficiency 10\%).  Hence, the $10^{4-5}~{\rm M_\odot}$ BHs, produced
through the primordial magnetic field, can indeed grow into the $\gsim
10^9~{\rm M_\odot}$ SMBHs by $z=6$.

The smallest total (dark matter+gas) mass that can collapse for $B_0
\simeq 4 \, \rm nG$ is $M \simeq 3 \times 10^{10} M_\odot$ (as
mentioned in section~3 above). These halos could form either as a
result of gravitational instability in the standard $\Lambda$CDM
model, or via PMF--induced density perturbations (for details on the
latter scenario, see e.g. SB08).  The abundance of halos in the
PMF--induced structure formation case drops very sharply for masses
above the Jeans mass (e.g. SS05), and for simplicity, we
conservatively drop this contribution in our analysis.  In the usual
$\Lambda$CDM model, using the fitting formula for the halo mass
function from Jenkins et al. (2001), and the current best--fit
cosmological parameters from Komatsu et al. (2009), we find that the
abundance of all $M>3\times 10^{10}~{\rm M_\odot}$ 
 halos at $z=10$ is $\approx5\times10^{-5}~{\rm
(comoving)~Mpc}^{-3}$.  At
the somewhat higher redshift of $z=15$, the abundance of halos above
the same mass drops sharply to $\approx3\times10^{-8}~{\rm cMpc}^{-3}$.

The space density of $\gsim 10^9~{\rm M_\odot}$ SMBHs, inferred from
the observed abundance of bright $z\approx6$ quasars, is $\sim
\epsilon_{\rm Q}^{-1}$ (comoving) Gpc$^{-3}$. Here $\epsilon_{\rm Q}$
denotes the duty cycle, defined as the fraction of the Hubble time
that $z=6$ supermassive black holes are observable as luminous
quasars.  Assuming a quasar lifetime of $\sim 50$ Myr (e.g. Martini
2004), we have $\epsilon_{\rm Q}\sim 0.05$, and the space density of
$z=6$ SMBHs is $\sim 20$ Gpc$^{-3}$.  Therefore, at redshift $z=10$, a
fraction as low as $f\sim (20~{\rm
Gpc}^{-3})/(5\times10^{-5}~{\rm
Mpc}^{-3})=3\times 10^{-3}$ of the whole population of
$M\gsim3\times 10^{10}~{\rm M_\odot}$ halos is sufficient to
account for the presence of these rare $z\approx6$ SMBHs.  Such a
small fraction could would correspond to $\sim 2.8\sigma$ upward
fluctuations of a Gaussian random PMF.  Because of the strong
reduction of the relatively high--mass halos at higher redshift,
essentially every $M\gsim 3\times 10^{10}~{\rm M_\odot}$ halo would
have to host SMBHs at $z\approx15$, to match the comoving abundance of
$z=6$ quasar BHs.  $z\approx 15$ is therefore the earliest epoch for
forming the heavy SMBH seeds as envisioned here.  As long as the PMF
amplitude is $ B\gsim 1.2 \, \rm nG$, the critical flux we find, $
B\approx 3.6 \, \rm nG$, would be reached by the $\sim3\sigma$ upward
fluctuations.

As mentioned above, our results are analogous to earlier work,
proposing that a UV flux keeps the gas hot, and results in SMBH
formation.  However, it is worth emphasizing two important differences
in these two scenarios. First, as Figure~1 shows, even though there is
a clear critical $B$--field value, dividing the thermal evolution into
two different regimes, even for $B$--field strengths below this value,
the minimum temperature reached by the gas is significantly elevated.
Hence, sub--critical magnetic fields can still significantly increase
the mass of the central object.  Second, the gas in the halos must
remain essentially free of any metals and dust, in order to avoid
fragmentation.  This may well be impossible to reconcile with the
large required flux, which has to arise from a close neighbor galaxy,
in the UV--irradiation scenario. On the other hand, in the
magnetic--heating case, there is a-priori reason why a region where
the PMF has an upward fluctuation would be more likely to be metal
enriched.  In fact, quite the contrary: the large magnetic Jeans mass
will suppress the collapse of gas into lower--mass halos at high
redshift, naturally excluding any prior metal--pollution farther up in
the merger tree of the halo in which the putative SMBH forms.

\section{Conclusions}

We found a plausible novel mechanism to form high-redshift SMBHs by
direct gas collapse in early dark matter halos, aided by heating from
the dissipation of a primordial magnetic field.  The model avoids many
of the assumptions required in earlier models (such as an extremely
high UV flux and the absence of ${\rm H_2}$ and of other molecules and
metals), but it does require a large primordial magnetic field, and
relies on metal--free primordial gas.  We expect that, in general,
{\em any other} heating mechanism, which can compete with atomic HI
cooling in the collapsing halo, down to a density of $n\sim
10^{3}~{\rm cm^{-3}}$, would produce the same effect as the $B$--field
utilized here.

Interestingly, our model requires a magnetic field strength that is
close to the existing observational upper limits.  We therefore expect
that the values $B_0 \simeq \hbox{a few} \times 10^{-9} \, \rm G$,
required to form SMBHs, could be detectable by the future CMB
experiments.  The upward fluctuations of such a strong primordial
magnetic field might also leave their trace on cosmic structure
formation: because of the high magnetic Jeans mass in these regions,
the formation of dwarf--galaxy sized halos, with masses below $M
\simeq \hbox{a few} \times 10^{10} M_\odot$, will be prevented.  We
expect this can lead to further constraints on our SMBH--formation
scenario, analogous to constraints on warm dark matter models (which
produce a similar suppression of small-scale structures; Barkana,
Haiman \& Ostriker 2001).

\section*{Acknowledgments}

ZH thanks the Raman Institute for their hospitality, where this work
was initiated, and the American Physical Society for travel
support. We also thank Biman Nath for many fruitful discussions.




\begin{figure}
\vspace{24pt} 
\includegraphics[width=0.8\textwidth]{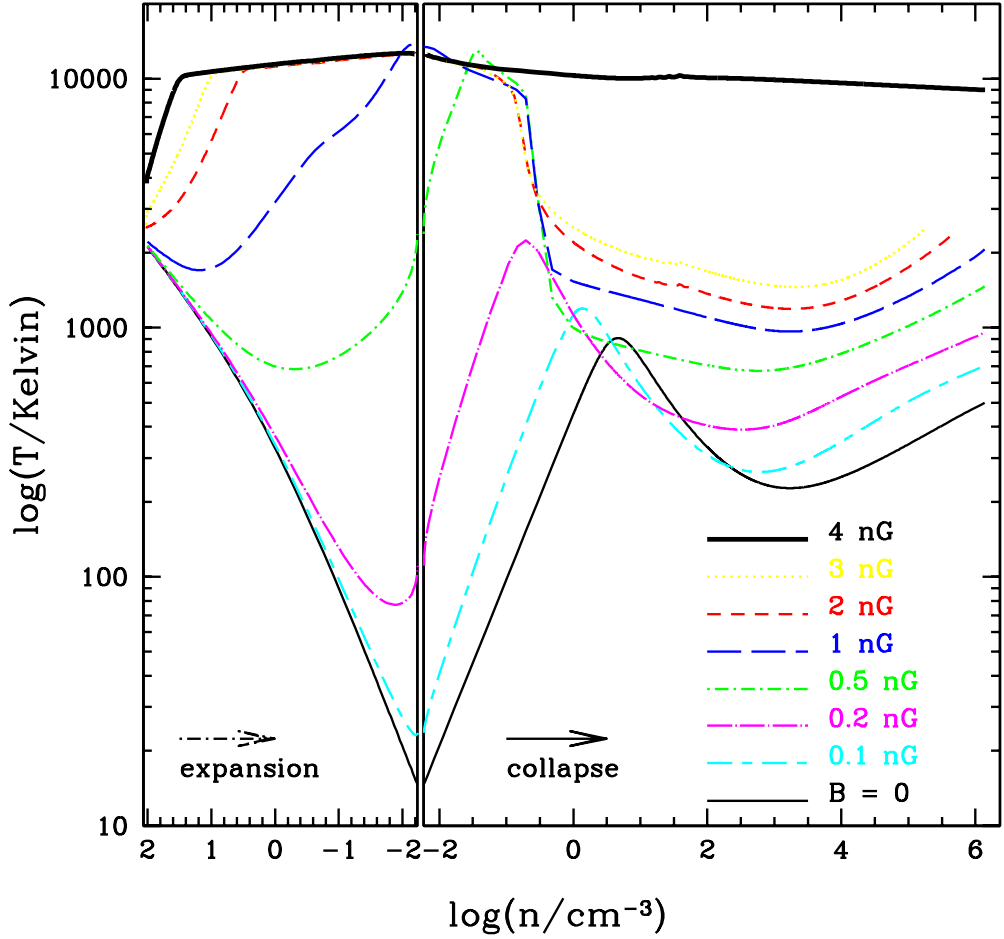} 
\caption{The temperature evolution of a patch of the intergalactic
  medium is shown as it initially expands and then turns around and
  collapses to high density.  The different curves correspond to
  different values of the assumed primordial magnetic field, as
  labeled.  The gas evolves from the left to the right on this figure.
  The left panel shows the expanding phase, starting from an initial
  density of $\approx 100~{\rm cm^{-3}}$ (corresponding to the mean
  density at redshift $z\simeq 800$) and ending at the turnaround just
  below $n=10^{-2}~{\rm cm^{-3}}$. The right panel follows the
  subsequent temperature evolution in the collapsing phase. }
\label{fig:nT}
\end{figure}

\clearpage
\newpage

\begin{figure}
\vspace{24pt}
\includegraphics[width=0.8\textwidth]{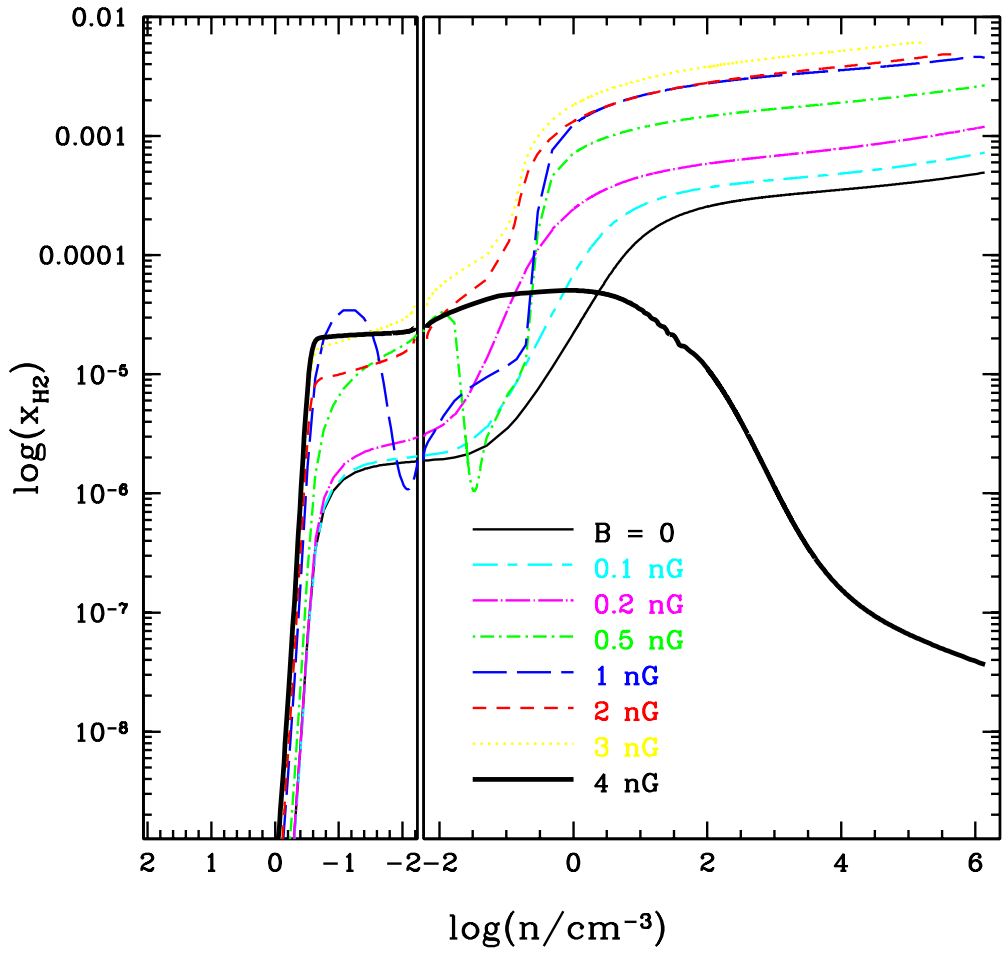}  
\caption{The evolution of the $\rm H_2$ fraction in the same gas
  clouds shown in Figure~\ref{fig:nT}.}
\label{fig:H2}
\end{figure}

\clearpage
\newpage

\begin{figure}
\vspace{24pt}
\includegraphics[width=0.8\textwidth]{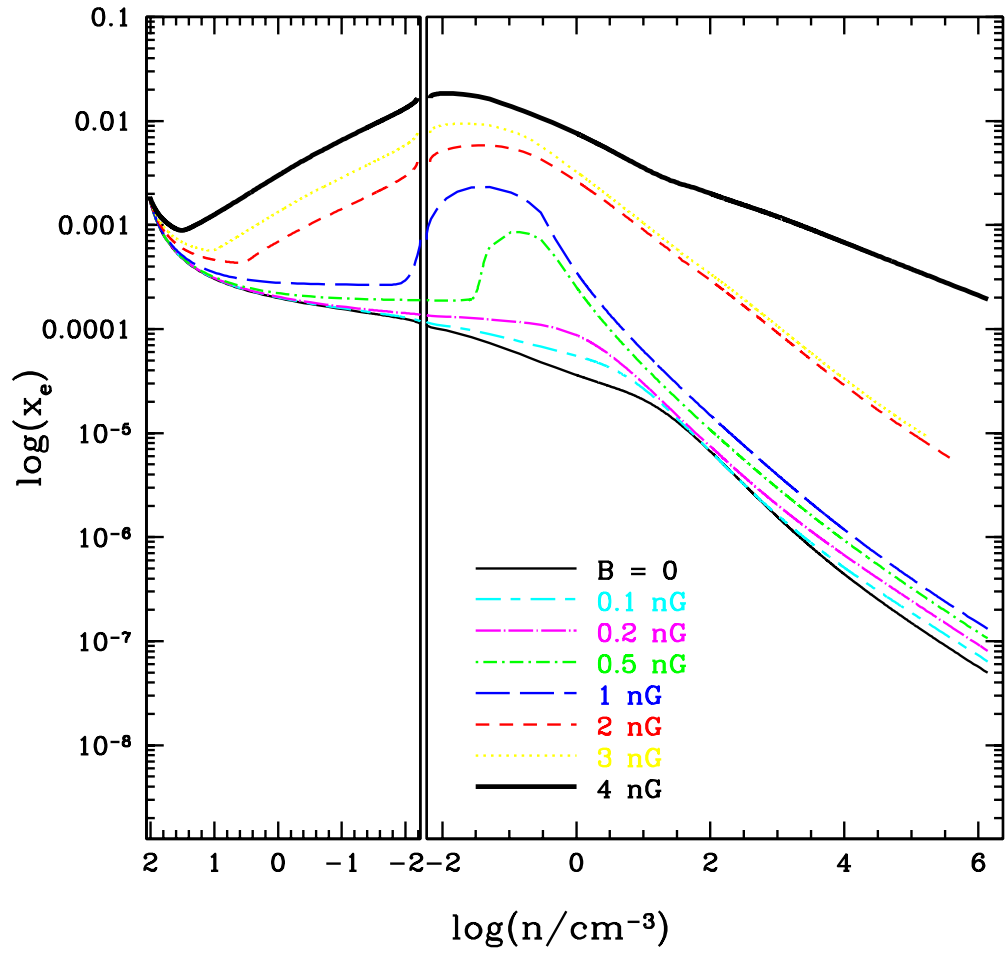}  
\caption{The evolution of the ionized fraction in the same gas clouds
  shown in Figure~\ref{fig:nT}.}
\label{fig:xe}
\end{figure}

\clearpage
\newpage

\begin{figure}
\vspace{24pt}
\includegraphics[width=0.8\textwidth]{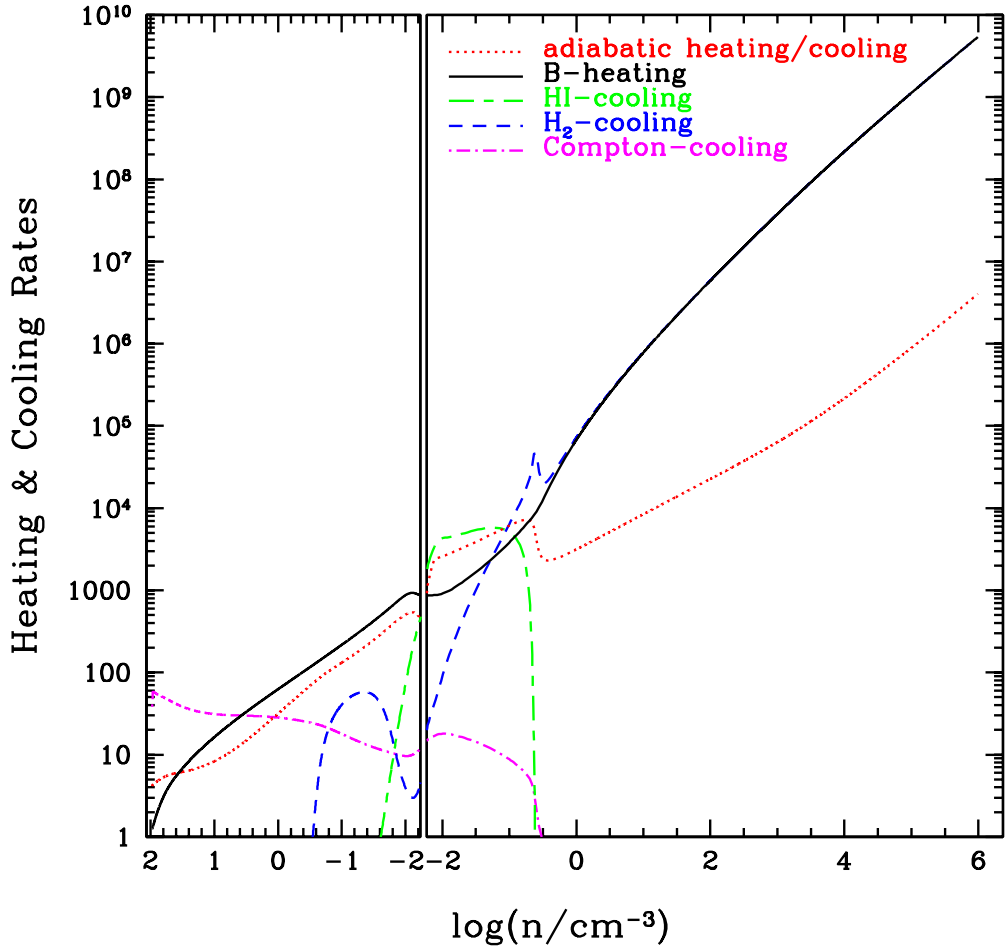}  
\caption{The heating and cooling rates are shown for various processes
as labeled, for $B=1$ nG. The rates are in the units $dt/dz = H_0^{-1}\Omega_m^{-1/2}(1+z)^{-5/2}$.}
\label{fig:rates1}
\end{figure}

\begin{figure}
\vspace{24pt}
\includegraphics[width=0.8\textwidth]{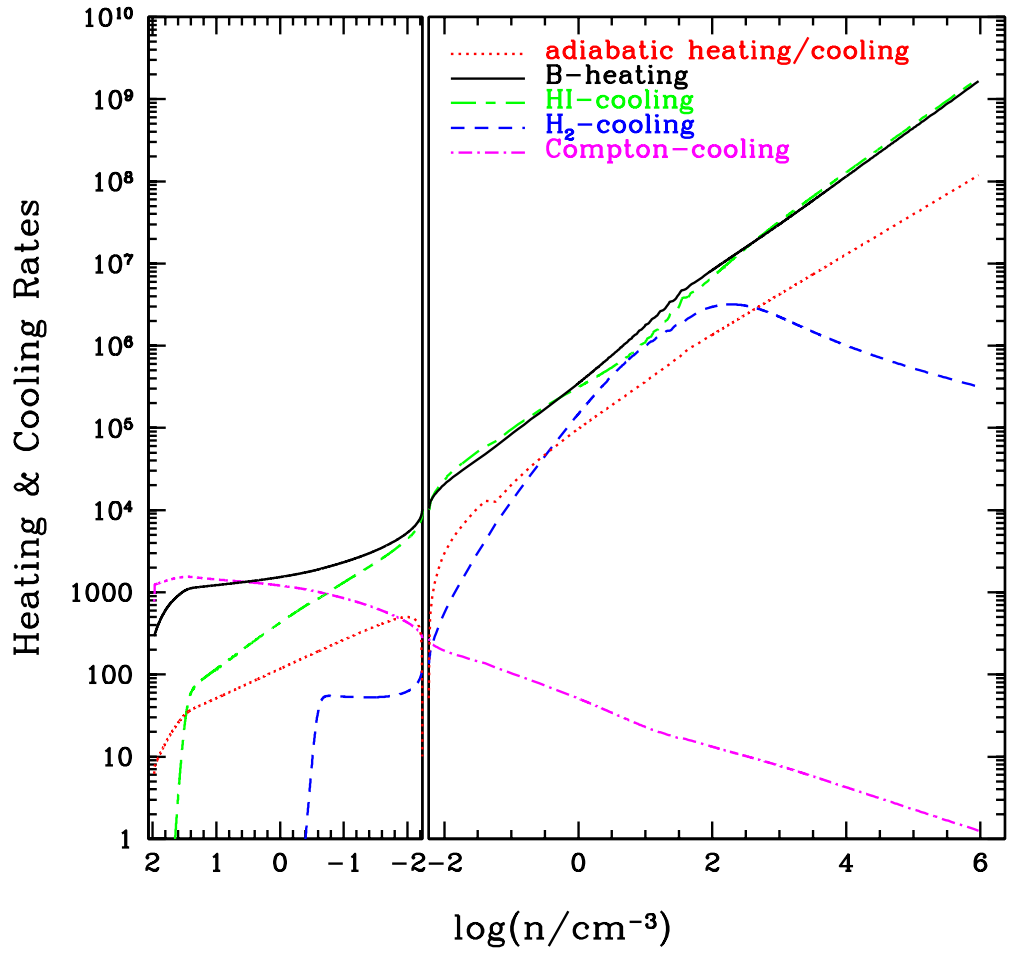}  
\caption{Same as Figure~4, but for $B=4$ nG.}
\label{fig:rates4}
\end{figure}


\begin{thebibliography}{4}

\bibitem[foo(2010)]{abel02} Abel, T., Bryan, G. L., Norman, M. L. 2002, Science, 295, 93

\bibitem[foo(2010)]{bho01} Barkana, R., Haiman, Z., \& Ostriker, J. P. 2001, ApJ, 558, 482

\bibitem[foo(2010)]{begelman08} Begelman, M., Rossi, E. M., \& Armitage, P. J. 2008, MNRAS, 387, 1649

\bibitem[foo(2010)]{begelman06} Begelman, M., Volonteri, M., \& Rees, M. J. 2006, MNRAS, 370, 289

\bibitem[foo(2010)]{bromm02} Bromm, V., Coppi, P. S., Larson, R. B. 2002, ApJ, 564, 23

\bibitem[foo(2010)]{bromm03} Bromm, V., \& Loeb, A. 2003, ApJ, 596, 34 

\bibitem[foo(2010)]{capitelli07} Capitelli, M., Coppola, C. M., Diomede, P. \& Longo, S. 2007, A\&A, 470, 811

\bibitem[foo(2010)]{cowling56} Cowling, T. G. 1956, MNRAS, 116, 114

\bibitem[foo(2010)]{db06} Dekel, A. \& Birnboim, Y. 2006, MNRAS, 368, 2 

\bibitem[foo(2010)]{dijkstra08} Dijkstra, M., Haiman, Z., Mesinger, A., \& Wyithe, S. 2008, MNRAS, 391, 1961

\bibitem[foo(2010)]{durrer} Durrer, R., Ferreira P. G. \&   Kahniashvili, T. 2000, Phys. Rev. D. , 61, 043001 

\bibitem[foo(2010)]{fan06} Fan, X. 2006 NewA Rev., 50, 665

\bibitem[foo(2010)]{finelli} Finelli, F., Paci, F. \&   Paoletti, D. 2008,  Phys.Rev.D, 78,023510

\bibitem[foo(2010)]{galli98} Galli, D., \& Palla, F. 1998, A\&A, 335, 403

\bibitem[foo(2010)]{giova08}  Giovannini, M. \& Kunze, K. E. 2008,  Phys. Rev. D., 77, 063003

\bibitem[foo(2010)]{gs05} Gopal \& Sethi 2005, 2005, Phys. Rev. D, 72,3003

\bibitem[foo(2010)]{gs03} Gopal, R. \& Sethi, S. K., 2003, Journal of Astrophysics and Astronomy, 24, 51

\bibitem[foo(2010)]{heger03} Heger, A., et al. 2003, ApJ, 591, 288

\bibitem[foo(2010)]{jenkins01} Jenkins, A. et al. 2001, MNRAS, 321, 372

\bibitem[foo(2010)]{kah05}  Kahniashvili, T.  \&   Ratra, B. 2005,  Phys. Rev. D, 71, 103006

\bibitem[foo(2010)]{kim96} Kim, E., Olinto, A.~V., \& Rosner, R.\ 1996, ApJ, 468, 28

\bibitem[foo(2010)]{komatsu09} Komatsu et al. 2009, ApJS, 180, 330

\bibitem[foo(2010)]{koushiappas04} Koushiappas, S. M., Bullock, J. S., \& Dekel, A. 2004, MNRAS, 354, 292

\bibitem[foo(2010)]{lewis} Lewis, A. 2004, Phys. Rev. D. 70, 43011 

\bibitem[foo(2010)]{li07} Li, Y., et al. 2007, ApJ, 665, 187

\bibitem[foo(2010)]{lodato06} Lodato, G., \& Natarajan, P. 2006, MNRAS, 371, 1813

\bibitem[foo(2010)]{mack} Mack, A.,  Kahniashvili, T. \&  A.  Kosowsky 2002, Phys. Rev. D 65, 123004 

\bibitem[foo(2010)]{martin96} Martin, P. G., Schwarz, D. H., \&  Mandy, M. E. 1996, ApJ, 461, 265

\bibitem[foo(2010)]{Martini04}  Martini, P.\ 2004 in ``Co-evolution of Black Holes and Galaxies'', Carnegie Observatories Astrophysics Series, Vol. 1, Ed. L. C. Ho. (Cambridge, U.K.: Cambridge University Press), p. 169

\bibitem[foo(2010)]{oh02} Oh, S. P., \& Haiman, Z. 2002, ApJ, 569, 558 

\bibitem[foo(2010)]{oh07} O'Shea, B.W., Norman, M. L., 2007, ApJ 654, 66

\bibitem[foo(2010)]{omukai01} Omukai, K. 2001, ApJ, 546, 635 (OM01)

\bibitem[foo(2010)]{omukai08} Omukai, K., Schneider, R., \& Haiman, Z. 2008, ApJ, 686, 801

\bibitem[foo(2010)]{regan09b} Regan, J. A., \& Haehnelt, M. G. 2009, MNRAS, 396, 343

\bibitem[foo(2010)]{schaerer02} Schaerer, D. 2002, A\&A, 382, 28

\bibitem[foo(2010)]{seshadri} Seshadri, T. R.  \&  Subramanian, K. 2001,  PRL, 87, 101301

\bibitem[foo(2010)]{sethi08} Sethi, S. K., Nath, B. B., \& Subramanian, K. 2008, MNRAS, 387, 1589 [S08]

\bibitem[foo(2010)]{sethi05} Sethi, S. K., \& Subramanian, K. 2005, MNRAS, 356, 778 [SS05]

\bibitem[foo(2010)]{schleicher09} Schleicher, D. R. G. et al. 2009, ApJ, 703, 1096

\bibitem[foo(2010)]{shang10} Shang, C., Bryan, G. L., \& Haiman, Z. 2010, MNRAS, 402, 1249

\bibitem[foo(2010)]{shu92} Shu, F. H. 1992, Gas Dynamics, University Science Books

\bibitem[foo(2010)]{spaans06} Spaans, M., \& Silk, J. 2006, ApJ, 652, 902

\bibitem[foo(2010)]{kandu98a} Subramanian, K.~\& Barrow, J.~D.\ 1998a, Phys. Rev. D, 58, 83502

\bibitem[foo(2010)]{subramanian3} Subramanian, K. \&  Barrow, J. D. 1998b, PRL, 81, 3575

\bibitem[foo(2010)]{subramanian4}  Subramanian, K.  \& Barrow, J. D. 2002, MNRAS, 335, L57

\bibitem[foo(2010)]{taka09}  Tanaka, T., \& Haiman, Z. 2009, ApJ, 696, 1798

\bibitem[foo(2010)]{turk09} Turk, M. J., Abel, T., \& O'Shea, B., 2009, Science, 325, 601

\bibitem[foo(2010)]{volonteri05} Volonteri, M., \& Rees, M. J. 2005, ApJ, 633, 624

\bibitem[foo(2010)]{volonteri06} Volonteri, M., \& Rees, M. J. 2006, ApJ, 650, 669

\bibitem[foo(2010)]{volonteri08} Volonteri, M., Lodato, G. \& Natarajan, P. 2008, MNRAS, 383, 1079

\bibitem[foo(2010)]{wasserman78} Wasserman, I.\ 1978, ApJ, 224, 337

\bibitem[foo(2010)]{widrow02} Widrow, L.~M.\ 2002, Reviews of Modern Physics, vol.~74, Issue 3, pp.~775-823, 74, 775 

\bibitem[foo(2010)]{wise08a} Wise, J.H., Abel, T. 2008, ApJ, 682, 745

\bibitem[foo(2010)]{yamazaki06} Yamazaki, D. G., Ichiki, K., Kajino, T., \& Mathews, G. J. 2006, ApJ, 646, 719

\bibitem[foo(2010)]{yamazaki08} Yamazaki, D. G.,  Ichiki, K.,  Kajino, T.  \&   Mathews, G. J. 2008,  Phys. Rev. D., 77, 043005

\bibitem[foo(2010)]{yamazaki10} Yamazaki, D. G., Ichiki, K., Kajino, T., \& Mathews, G. J. 2010, Phys. Rev. D, 81, 023008

\bibitem[foo(2010)]{yoshida08} Yoshida, N., Omukai, K., \& Hernquist, L. 2008, Science, 321, 669

\end{thebibliography}
\end{document}